\documentclass[%
 aip,
 jmp,%
 amsmath,amssymb,
 reprint,%
]{revtex4-1}

\usepackage{graphicx}
\usepackage{epstopdf}
\usepackage{dcolumn}
\usepackage{bm}

\begin{document}

\preprint{AIP/123-QED}

\title[Magnetic turbulence: soft-hard MHD regimes ]{Magnetic turbulence and pressure gradient feedback effect of the $1/2$ mode soft-hard MHD limit in Large Helical Device}

\author{J. Varela}
 \email{jvrodrig@fis.uc3m.es}
\affiliation{LESIA, Observatoire de Paris, CNRS, UPMC, Universite Paris-Diderot, 5
place Jules Janssen, 92195 Meudon, France}
\author{K.Y. Watanabe}
\affiliation{National Institute for Fusion Science, Oroshi-cho 322-6, Toki 509-5292, Japan}
\author{S. Ohdachi}
\affiliation{National Institute for Fusion Science, Oroshi-cho 322-6, Toki 509-5292, Japan}
\author{Y. Narushima}
\affiliation{National Institute for Fusion Science, Oroshi-cho 322-6, Toki 509-5292, Japan}

\date{\today}

\begin{abstract}

The aim of this study is to analyze the feedback process between the magnetic turbulence and the pressure gradients in LHD inward-shifted configurations as well as its role in the transition between the soft-hard MHD regimes for instabilities driven by the mode $1/2$ in the middle plasma. In the present paper we summarize the results of two simulations with different Lundquist numbers, $S = 2.5 \times 10^5$ and $10^6$, assuming a plasma in the slow reconnection regime. The results for the high Lundquist number simulation shows that the magnetic turbulence and the pressure gradient in the middle plasma region of LHD are below the critical value to drive the transition to the hard MHD regime, therefore only relaxations in the soft MHD limit are triggered ($1/2$ sawtooth like events) [Phys. Plasmas 19, 082512 (2012)]. In the case of the simulation with low Lundquist number, the system reaches the hard MHD limit and a plasma collapse is observed.

\end{abstract}

\maketitle

\section{Introduction \label{sec:introduction}}

Some linear simulations in Large Helical Device (LHD) predicte a strong interchange activity in the middle plasma for high beta inward-shifted configurations \cite{1,2,3}, but there is not experimental evidences of large instabilities, only minor relaxations like the $1/2$ sawtooth like events \cite{4,5}. There is a stabilizing mechanism avoiding the explosive growth rate of the interchange modes \cite{6,7}. If a mode is unstable the pressure profile is flattened around the rational surfaces and the mode growth saturates, leading to a staircase-like pressure profile. In other studies it was reported the possibility of a disruptive behavior of the plasma if the interaction between modes with different helicity is strong \cite{8}. It is important to find out the mechanism that regulates the efficient of the mode helical coupling because it can be related with the transition between the plasma soft/hard MHD regimes. 

The MHD instabilities in LHD inward configuration can be driven in the hard MHD limit or in the soft MHD limit \cite{9,10}. The effect of a relaxation in the soft MHD limit is less dangerous than in the hard MHD limit, because the drop of the device efficiency to confine the plasma is smaller. The instabilities driven in the soft MHD limit are local, only a small plasma region is affected, and the unstable modes saturate as fast as the pressure profile is flattened around the unstable rational surface. In the case of the hard MHD limit, the instability spreads on modes with different helicities along the plasma and the events driven are global relaxations. The events studied in the soft MHD limit are the non resonant $1/3$ and $1/2$ sawtooth like events, and the internal disruptions and resonant $1/3$ sawtooth like events in the hard MHD limit \cite{9,11,12}. There is experimental evidence of these events in LHD, expect in the case of the internal disruption \cite{13}. 

We propose in this paper a feedback process between the magnetic turbulence and the pressure gradients like a candidate to explain the transition from the soft to the hard MHD limit for a plasma in the middle region of the LHD device, where the destabilizing effect of the mode $1/2$ dominates \cite{14,15}. This feedback effect is based in the properties of the magnetic reconections that take place in the simulation, like the reconnection rate and their efficiency to generate magnetic turbulence \cite{16,17}.   

The magnetic reconnections are a source of MHD turbulence because it increases the Ohmic and viscous dissipation by local current and vorticity structures \cite{18}. The reconnected islands produce a flux rearrangement that reinforces the magnetic turbulence. The magnetic turbulence perturbates the flux and magnetic surfaces of the plasma enhacing the pressure gradients and the destabilizing effect of the unstable modes. The plasma reaches the hard MHD regime if the magnetic turbulence and the pressure gradients are large enough to sustain a strong mode coupling and a large magnetic islands overlapping \cite{19,20}. Large stochastic regions appear in the plasma reducing the confinement capacity of the device \cite{21}, breaking down the magnetic surfaces and producing a large amount of magnetic turbulence.

In the present research we study the feedback process between the magnetic turbulence and the pressure gradients in two simulations with different Lundquist numbers ($S$), a high $S$ case with $10^6$ and a low $S$ case with $2.5 \times 10^5$. Each simulation describes a plasma with different stability properties due to the effect of the magnetic reconnections and the feedback process in the system evolution.  The Lundquist number is defined like the ratio between the Alfv\'en crossing time scale and the resistive diffusion time scale. We assume a plasma in the slow reconnection regime, Sweet-Parker theory \cite{22}, not in the fast reconnection regime of collisionless plasmas \cite{23}.

The simulation with a low Lundquist number is expected to be more unstable as it was observed in reversed-field pinches \cite{24,25} and Tokamaks \cite{26}, where the magnetic turbulence and the pressure gradient decrease if the device operates in a regime with a high Lundquist number (high conducting plasmas). 

We made the simulations using the code FAR3D \cite{27,28,29}. The code solves the reduced non-linear resistive MHD equations evolution from a perturbation driven on a VMEC equilibria \cite{30}, reconstructed from the measurements of the electron density and the temperature using the Thomson scattering and the electron cyclotron emission data. 

The paper is organized as follows. Sec. II, numerical model, equilibrium and simulation parameters. Sec. III, linear study.  Sec. IV, comparative study.  Sec. V, analysis of the simulations main events. VI, discussion and conclusions.

\section{Numerical model \label{sec:model}}

The reduced equations in FAR3D code use the assumptions of high-aspect ratio and medium beta values (of the order of the inverse aspect ratio $\varepsilon$). There is no averaging in the toroidal angle so we use in the simulations an exact three-dimensional equilibrium. We include in the calculations the linear coupling of the toroidal modes families. 

The FAR3D uses the finite differences in the radial direction and a Fourier expansion in the angular coordinates. It is a semi-implicit numerical scheme in the linear terms and explicit in the non linear terms. The method accuracy is ($\Delta t^2$).

The system geometry is described in Boozer coordinates \cite{31}. The equilibrium flux coordinates are ($\rho$, $\theta$, $\zeta$) with $\sqrt g$ the Jacobian of the coordinate transformation. The generalized radial coordinate $\rho$ is  proportional to the square root of the toroidal flux function and normalized to the unity in the edge. The poloidal angle is $\theta$  and the toroidal angle is $\zeta$.

The velocity, magnetic field and vorticity ($U$) can be expressed like:

\begin{equation}
 \mathbf{v} = \sqrt{g} R_0 \nabla \zeta \times \nabla \Phi
\end{equation}

\begin{equation}
\mathbf{B} = R_0 \nabla \zeta \times \nabla \psi
\end{equation}

\begin{equation}
\mathbf{U} =  \sqrt g \left[{ \nabla  \times \left( {\rho _m \sqrt g {\bf{v}}} \right) }\right]^\zeta
\end{equation}
The stream function $\Phi$ is  proportional to the electrostatic potential, $\psi$ is the perturbation of the poloidal flux and $\rho_{m}$ is the mass density. The eigenfunctions have an equilibrium and a perturbative term. The code solves the time evolution of the perturbed poloidal flux, vorticity and pressure.

The dimensionless form of the equations are:

\begin{equation}
\frac{{\partial \psi }}{{\partial t}} = \nabla _\parallel  \Phi  + \frac{\eta}{S} J_\zeta
\end{equation}
\begin{eqnarray} 
\frac{{\partial U}}{{\partial t}} = - {\mathbf{v}} \cdot \nabla U + \frac{{\beta _0 }}{{2\varepsilon ^2 }}\left( {\frac{1}{\rho }\frac{{\partial \sqrt g }}{{\partial \theta }}\frac{{\partial p}}{{\partial \rho }} - \frac{{\partial \sqrt g }}{{\partial \rho }}\frac{1}{\rho }\frac{{\partial p}}{{\partial \theta }}} \right) \nonumber\\
 + \nabla _\parallel  J^\zeta  + \mu \nabla _ \bot ^2U
\end{eqnarray} 
\begin{equation}
\label{peq}
\frac{{\partial p}}{{\partial t}} =  - {\mathbf{v}} \cdot \nabla p + D \nabla _ \bot ^2p + Q
\end{equation}
The term $J_\zeta$ is the toroidal current density. The lengths are normalized to the minor radius ($a$) and the time to the poloidal Alfv\'en time ($\tau_{hp} = R_0 (\mu_0 \rho_m)^{1/2} / B_0$). The resistivity, the magnetic field and pressure are normalized to their values in the magnetic axis. The resistive time is defined like ($\tau_{R} = a^2 \mu_0 / \eta_0$). The operator ($\nabla_{\parallel}$) is the derivation in the direction parallel to the magnetic field:

\begin{equation*}
\nabla_{||} = \frac{\partial}{\partial\zeta} + \rlap{-} \iota\frac{\partial}{\partial\theta} - \frac{1}{\rho}\frac{\partial\tilde{\psi}}{\partial\theta}\frac{\partial}{\partial\rho} + \frac{\partial\tilde{\psi}}{\partial\rho}\frac{1}{\rho}\frac{\partial}{\partial\theta},
\end{equation*}
with $\rlap{-} \iota$ the rotational transform. The coefficient $D$ is the collisional cross-field transport, $\mu$ is the collisional viscosity for the perpendicular flow and $Q$ is the energy source term.

\subsection{Equilibrium}

The equilibrium is a free-boundary VMEC calculated before a sawtooth like event is driven\cite{13}. This equilibrium is obtained during a LHD operation in an inward-shifted configuration with the vacuum magnetic axis at $R_{0} = 3.6$ m. It is a high density plasma produced by pellet injection, strongly heated with three Neutral Beam Injectors, without net toroidal current, a magnetic field in the magnetic axis of $2.75$ T, an aspect ratio of $0.16$ and a $\beta$ value in the magnetic axis of $1.48$ $\%$ \cite{1}. Fig. 1 shows the pressure and rotational transform profiles.   

\begin{figure}[h]
\centering
\includegraphics[width=0.3\textwidth]{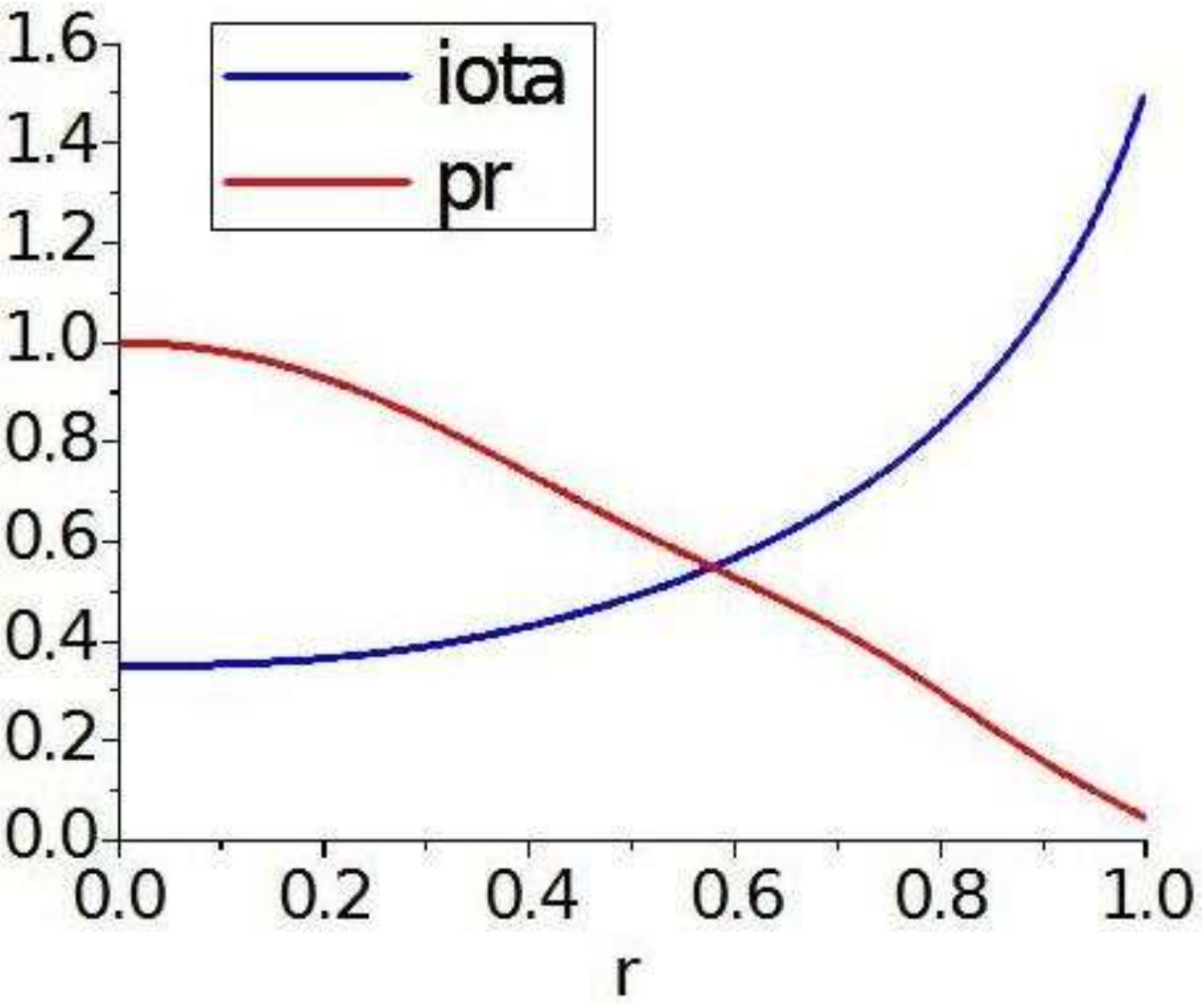}
\caption{Equilibrium rotational transform and pressure profiles.}
\end{figure}

\subsection{Simulation parameters}

The value of the dissipative terms $D = 1.25 \times 10^{-5}$ and $\mu = 7.5 \times 10^{-6}$ are chosen to damp the modes of toroidal families with $n > 10$ which are an energy sink in the simulation. The energy leak of the system is balanced with the energy source $Q$, a gaussian centered near the magnetic axis dynamically fitted to kept almost constant the integral area of the pressure. The $\beta_{0}$ value in the simulation is $1.184$ $\%$.

The simulation has an uniform radial grid of $500$ point and $515$ Fourier modes. The range of toroidal numbers is $n = [0,30]$. The $n = 0$ toroidal family has $m = [0,5]$ poloidal modes. The dynamic poloidal modes are chosen to improve the resolution between the magnetic axis and the middle plasma region. The linear simulation was tested with a grid radial resolutions of $1000$ and $2000$ points in the low and high $S$ cases. The growth rate and eigenfunctions shapes agree with the $500$ results.

We analyze the evolution of the equilibria for the Lundquist numbers $2.5 \times 10^5$ and $10^6$. To avoid strong overshots the Lundquist number is increased in $100$ steps. The simulation data affected by the $S$ value run up is not considered in the analysis. The Lundquist number is constant along the normalized minor radius. The simulation with $S = 10^6$ has a Lundquist number around $2$ orders smaller than the experimental value near the magnetic axis, almost the same in the middle plasma and one order higher than the outer plasma. The aim of the simulation is to study the stability properties of the plasma in the middle region where the destabilizing effect of the mode $1/2$ dominates, therefore the simulation with $S = 10^6$ is the best approximation of the real LHD plasma.  The simulations with lower $S$ value are useful to study the stability properties of the system, but they are not realistic conditions because the plasma is too resistive and the reconnection rate too fast.

The simulation remains in the slow reconnection regime defined by the Sweet-Parker theory if the Lundquist number is below a critical value. Above this limit the current sheet is unstable to a super-Alfv\'enic plasmoid instability and the system enters in a new nonlinear regime where fast reconnections are triggered, with a reconnection rate much faster than the Sweet-Parker theory \cite{32,33}. The single fluid description of the present study is not valid in the fast reconnection regime. In our case the reconnection can be driven only via the Ohm's term, but in collisionless plasmas it is negligible compared with the electron inertia, Hall effect and the electron viscosity terms. To reproduce the Physics in the fast reconnection regime the simulation should be done at least with a two fluid model \cite{17}. Recent studies suggest that the $\beta$ and plasma density have a first other effect in the transition between the regimes with slow and fast reconnections. A plasma with a low $\beta$ and high density remains in the slow reconnection regime for higher $S$ values \cite{34} because its collisionality increases \cite{35,36,37}. The present study simulations with low $S$ are below the critical value for the transition and the high $S$ simulations are marginal unstable to the plasmoid instability, therefore the single fluid model is valid in first approximation and we assume the Ohm's term like the dominant reconnection driver. The ion inertial length is 5 orders smaller than the device minor radius so the single MHD description is suitable.

\section{Linear study \label{sec:linear}}

We made several linear simulation to study the growth rate and the width of the pressure and the radial magnetic field eigenfunctions of the $1/2$ mode for different Lundquist numbers (fig. 2). The plasma in the high $S$ simulations behaves like an ideal plasma and the low $S$ plasmas are more resistive. The growth rate of the $1/2$ mode and the eigenfunctions width decreases as the $S$ value increases. The magnetic islands width is smaller in high $S$ plasmas so the magnetic islands overlapping is weaker and the plasma remains in the soft MHD regime. The hard MHD limit can be reached easily in the simulations with low $S$ because the eigenfunction width is larger and the mode growth rate is higher. 

\begin{figure}[h]
\centering
\includegraphics[width=0.5\textwidth]{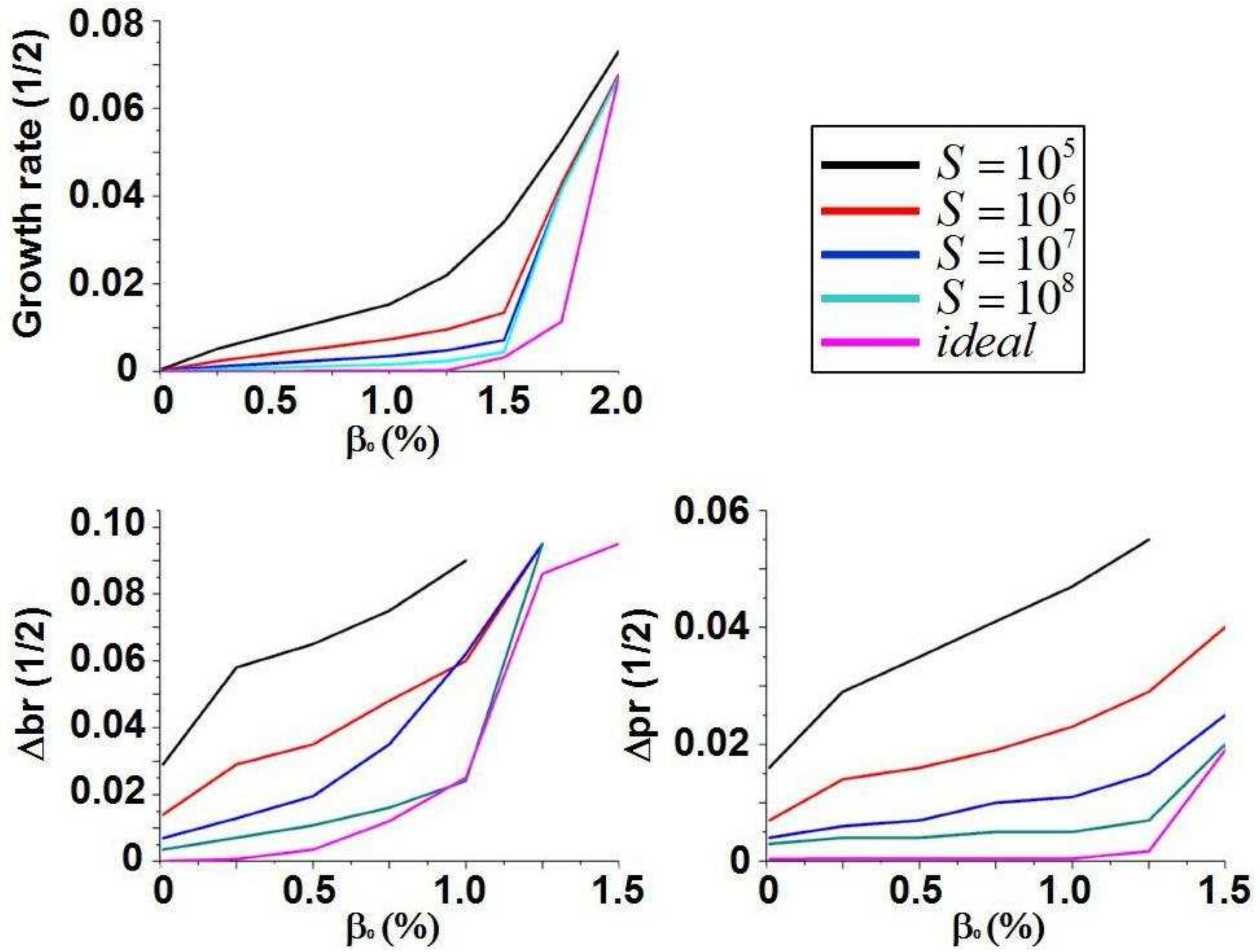}
\caption{Mode $1/2$ linear growth rate, pressure and radial magnetic field eigenfunctions widths for the ideal case and Lundquist numbers between $10^5$ to $10^8$.}
\end{figure}

The linear simulation that is closer to the experimental conditions in the middle plasma region of the LHD device is the case with $S = 10^{6}$. Compared with the $S = 10^{5}$ simulation, the growth rate for $\beta_{0} = 1$ $\%$ is less than half and the width of the radial magnetic field and pressure functions are almost half too. The simulations with $S = 10^{7}$ and $S = 10^{8}$ are close to an ideal plasma and they are only marginal unstable. The Physics of the soft-hard MHD transition in the plasma with low $S$ and high $S$ are different because the high $S$ plasmas are in the fast reconnection regime, while the low $S$ plasmas are in the slow reconnection regime. The high $S$ plasmas, $S > 10^6$, are out of the scope of this study.

If we compare these results with the experiment, the destabilizing effect of the mode $1/2$ triggers the $1/2$ sawtooth like activity in LHD inward-shifted configurations. This is accordance with the results of the simulation of $S = 10^{6}$, but not with the simulations for higher $S$ values. Another research concluded that in the LHD inward-shifted operations, if the $S$ value decreases and the $\beta$ gradient increases the $1/2$ sawtooth like activity is triggered more often \cite{1}. 

The next step of this research is to analyze if the conclusions in the linear approximation are valid in the non linear case. 

\section{Comparative study \label{sec:comparative}}

The comparative study of the normalized magnetic and kinetic energy (Fig. 3) shows a different evolution in the low $S$ and high $S$ simulations. The main events are driven at $t = 0.6233$ s (red arrow) in the low $S$ case and at $t = 0.6191$ s (blue arrow) in the high $S$ case like the dominant fluctuations in the profiles. The main events are out of phase and the profile structures are different.

\begin{figure}[h]
\centering
\includegraphics[width=0.35\textwidth]{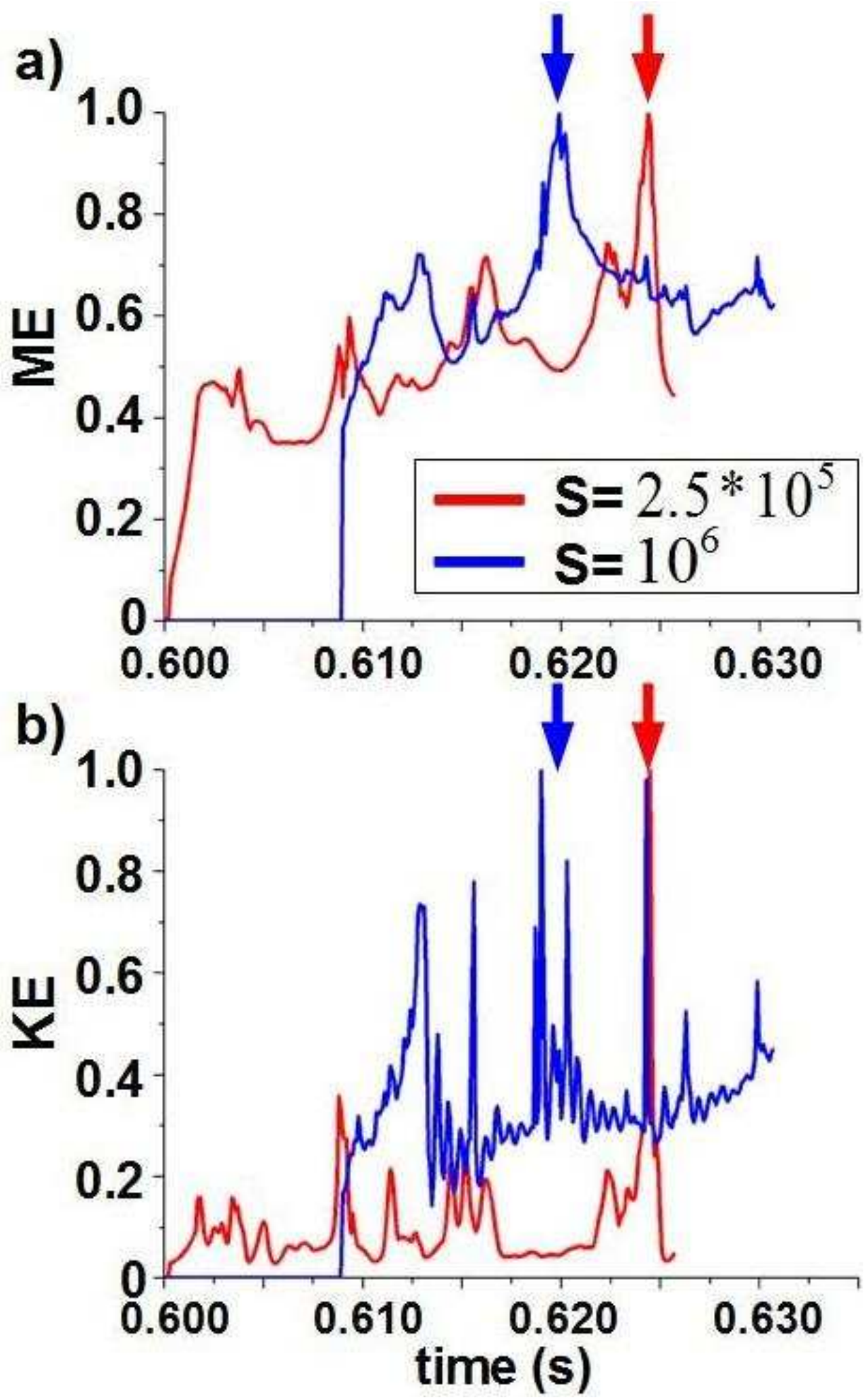}
\caption{Normalized magnetic (a) and kinetic energy (b) in simulations with the Lundquist numbers $2.5 \times 10^5$ and $10^6$. The arrows show the main events in each simulation.}
\end{figure}

If we compare the evolution of the magnetic energy of the single modes, Fig. 4, the modes in the inner plasma ($3/7$, $2/5$, $1/3$) and middle plasma ($1/2$) are more unstable in the low $S$ simulation. The modes in the outer plasma ($2/3$, $3/4$, $3/5$, $2/2$) shows a similar evolution in both cases. During the main event in the low $S$ case, Fig. 4a, there is a correlation between the modes ME oscillations and local maximums indicating that the system relaxation affects all the plasma regions. The modes $1/2$ and $3/5$ show the largest oscillations like a energy peak followed by a step drop, pointing out that the middle plasma is strongly destabilized. During the main event in the high $S$ simulation, Fig. 4 b, the modes in the inner plasma region are weakly destabilized compared with the low $S$ case, and the largest energy oscillations are observed for the modes $1/2$, $3/5$ and $3/4$, then the system relaxation is located between the middle and the outer plasma.

\begin{figure}[h]
\centering
\includegraphics[width=0.35\textwidth]{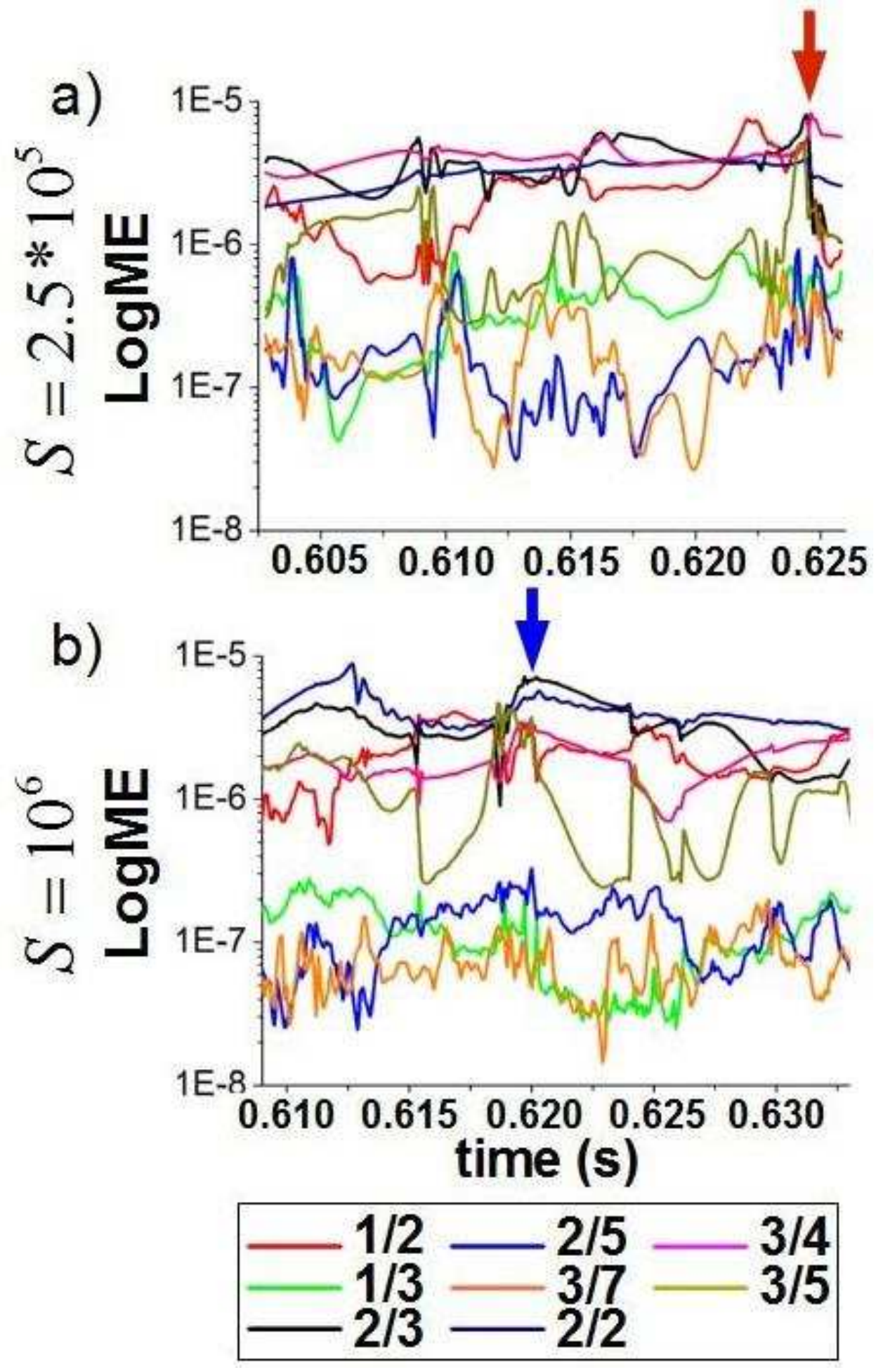}
\caption{Magnetic energy evolution of the single mode for the simulations with $S = 2.5 \times 10^5$ (a) and $10^6$ (b).}
\end{figure}

The evolution of the system energy loss, Fig. 5, defined like $E = \int pdV$ with $dV$ the differential element of the plasma volume, shows the energy leak of the system when a relaxation is triggered. A drop in the profile is linked with a deterioration of the device ability to confine the plasma. In the low $S$ simulation the successive system relaxation drive large energy leaks out of the system with a maximum during the main event. In the high $S$ case the system relaxations drive a local flattening or a small drop of the system energy, like in the main event, but the energy leaks are much smaller than in the low $S$ case. The plasma relaxations in the low $S$ simulation are stronger than in the high $S$ case. The main event of the low $S$ simulation is more dangerous for the device efficiency to confine the plasma because it drives a large energy loss, while the main event in the high $S$ simulation triggers only a small energy loss. The green line in Fig. 5 indicates the end of the data affected by the numerical overshot in the $S$ value run up and it is not analyzed.

\begin{figure}[h]
\centering
\includegraphics[width=0.4\textwidth]{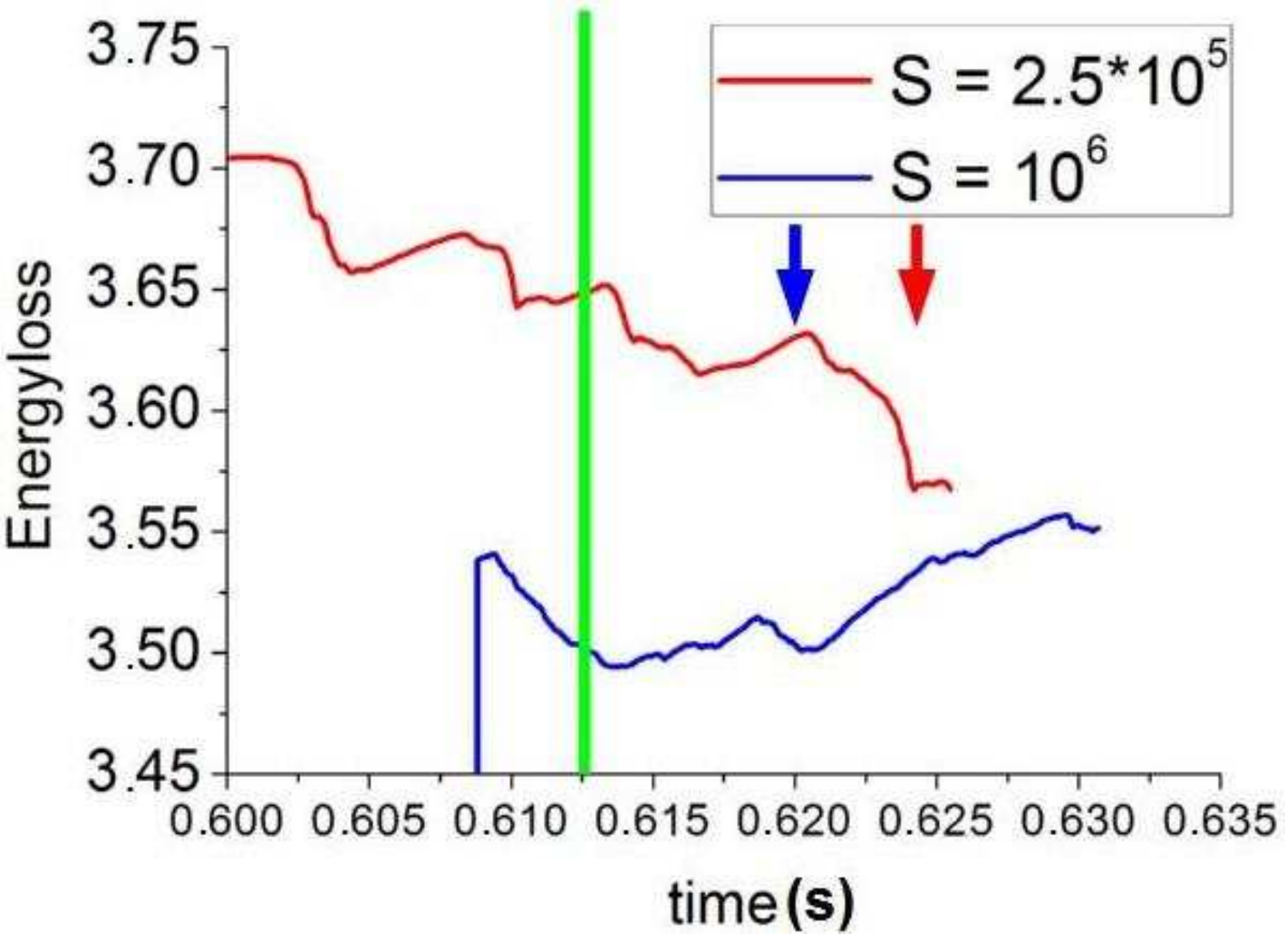}
\caption{Energy loss. The green line indicates the end of the overshot data region.}
\end{figure}

The pressure gradients, Fig. 6, in the low $S$ simulation are stronger than in high $S$ case. During the system relaxations in the low $S$ simulation there are large oscillations of the pressure gradient, specially in the inner and middle plasma region. The pressure gradient increases along the low $S$ simulation reaching a maximum during the main event before the profile drop in all the plasma regions. The pressure gradient fluctuations in the high $S$ simulation are weaker, they only affect a single plasma region and they don't increase along the simulation. The largest oscillations are driven during the main event like a peak in the profile mainly between $\rho = 0.3$ and $0.5$. During the low $S$ simulation the pressure gradient builds up until a critical value is reached and a strong system relaxation is triggered. This event affects all the plasma regions because there is a strong correlation of the pressure gradient between the inner plasma to the outer plasma. In the high $S$ case the pressure gradients don't build up along the simulation and the oscillations only affects a plasma region, therefore the relaxations are local. We show only the most representative graphs at the normalized radius points $0.1$ (almost the magnetic axis region), $0.3$ (inner plasmas), $0.5$ (middle plasma) and $0.7$ (outer plasma), but the pressure gradient oscillations are not local effects, they are extended along each plasma region.

\begin{figure}[h]
\centering
\includegraphics[width=0.5\textwidth]{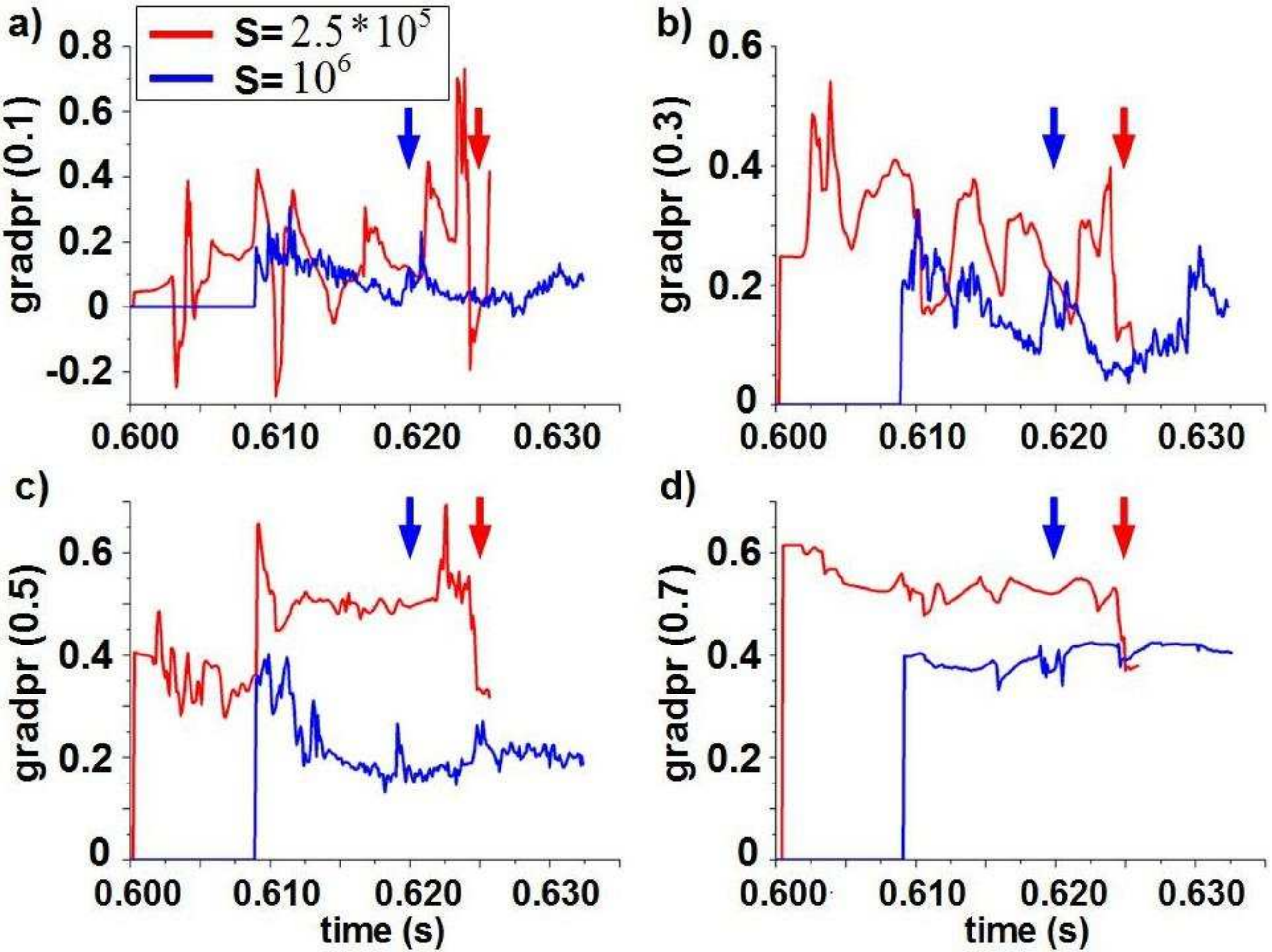}
\caption{Pressure gradient at the normalized minor radius $\rho = 0.1$ (a), $0.3$ (b), $0.5$ (c) and $0.7$ (d).}
\end{figure}

The magnetic turbulence is proportional to the perturbed magnetic field $\left | \widetilde{B} \right | / B_{0}$ , Fig. 7. The magnetic turbulence in the low $S$ case is larger than in the high $S$ case and its average slope is positive increasing during the simulation. In the high $S$ case the average slope is negative and the magnetic turbulence decreases along the simulation. The magnetic turbulence of a plasma in the slow reconnection regime should increase as the $S$ value decreases, because the magnetic reconnection rate increases. This is the case in the low $S$ simulation, the evolution of the magnetic turbulence shows an average positive slope, while in the high $S$ simulation the slope is negative. The peaks in the magnetic turbulence during the low $S$ cases are related with the overlapping of the magnetic islands during the system relaxations, that generates a large amount of turbulence in the regions with stochastic magnetic field, specially during the main event. 

\begin{figure}[h]
\centering
\includegraphics[width=0.4\textwidth]{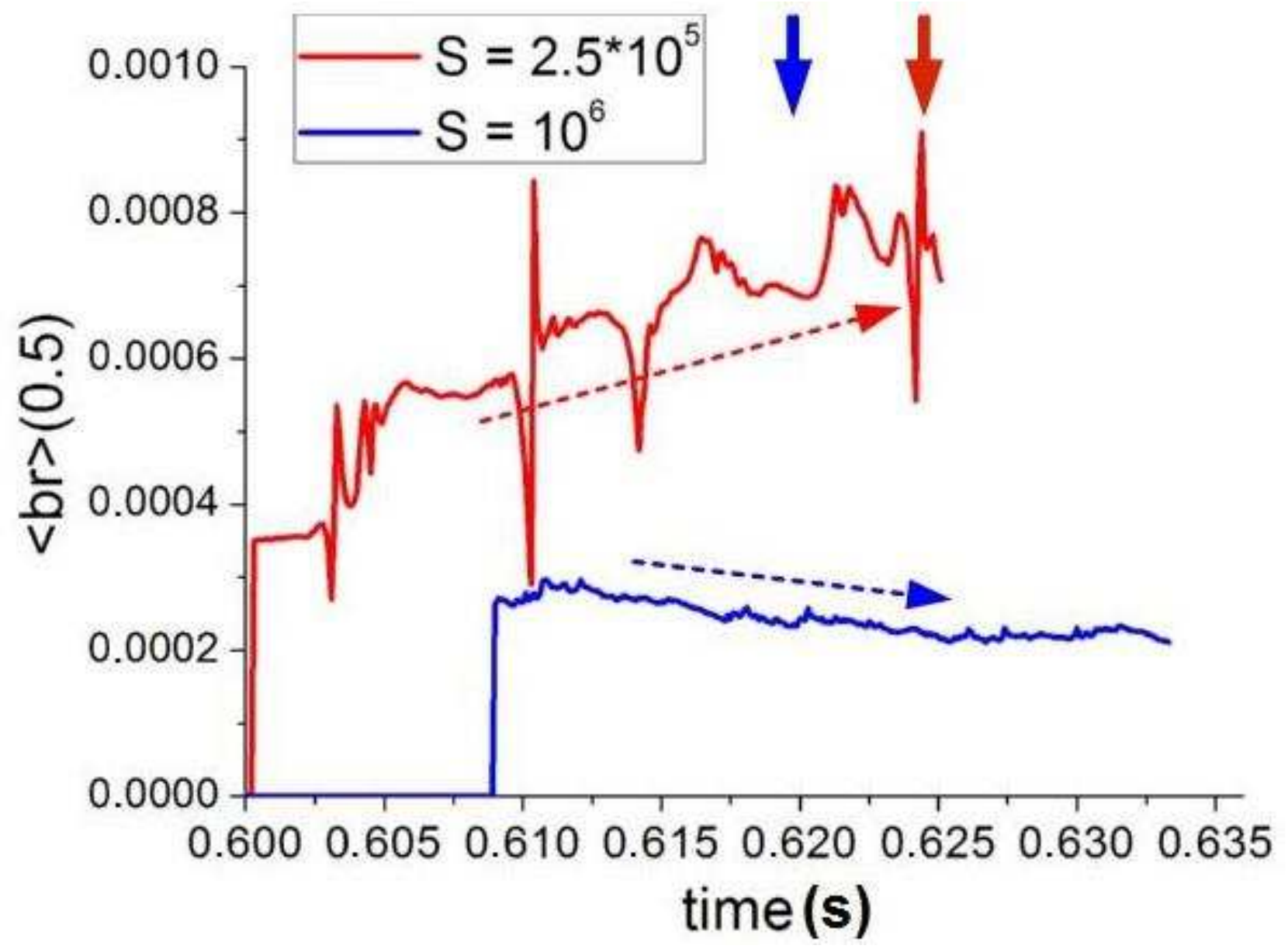}
\caption{$\left | \widetilde{B} \right | / B_{0}$ evolution at the normalized minor radius $\rho = 0.5$. Discontinuous lines are the averaged slopes of the graphs.}
\end{figure}

To support these results we analyze in the next section the characteristics of the main events driven in each simulation.

\section{Analysis of the simulations main events \label{sec:main}}

We use the next theoretical diagnostics to analyze the equilibria characteristics:

1) Averaged pressure profile; a pressure profile flattening or profile inversion shows the plasma region along the normalized minor radius where there is an unstable rational surface. It is expressed like $<p> = p_{eq}(\rho) + p'_{00}(\rho)$. The brackets means average over the flux surface,  $p_{eq}$ is the equilibrium pressure and $p'_{00}$ is the component $n = 0$ and $m = 0$ of the perturbed pressure.

2) Instantaneous profile of the rotational transform; it gives information of the radial location of the rational surfaces in the plasma. It is expressed like:

\begin{equation}
\label{iota}
\rlap{-} \iota (\rho)+ \tilde{\rlap{-} \iota}(\rho) = \rlap{-} \iota+ \frac{1}{\rho}\frac{\partial\tilde{\psi}}{\partial\rho}
\end{equation}

3) Two-dimensional contour plots of the pressure; it shows the shape of the flux surfaces. It is expressed like $p =  p_{eq}(\rho) + \sum_{nm} p'_{nm}(\rho) cos(m\theta + n\zeta)$.

4) Poincar\'e plots of the magnetic field; it presents the instantaneous topology of the magnetic field. There are two versions, the plots of family modes with single helicity and the plots with all the modes of the simulation. The single helicity plots give information of the magnetic field shape and the magnetic islands. The full modes plots exhibit the stochastic regions in the plasma related with the magnetic turbulence of the system. The Poincar\'e plots overestimate the size of the stochastic regions, the important information is the evolution of the stochastic regions during the events.

5) Magnetic potential; it shows the evolution of the magnetic well and magnetic hill regions in the plasma. The interchange modes are stable if the plasma is in a region with magnetic well, $\langle V(r) \rangle < 0 $, and unstable in the magnetic hill region, $\langle V(r) \rangle > 0 $.

The main event in the low $S$ simulation is driven at $t = 0.6233$ s. There are several deformations of the pressure profile in the inner, middle and outer plasma around $\rho = 0.25$ (mode $3/8$), $\rho = 0.4$ (mode $3/7$), $\rho = 0.5$ (mode $1/2$), $\rho = 0.65$ (mode $2/3$), but the destabilizing effect of the $1/2$ mode dominates, Fig. 8 a. At $t = 0.6238$ s the profile deformation in the inner plasma enhances, around $\rho = 0.25$. At $t = 0.6241$ s there is a profile inversion in the outer plasma around $\rho = 0.66$. At $t =  0.6241$ and $0.6250$ s there is a pressure profile flattening near the magnetic axis driven by the mode $1/3$. The mode $1/3$ enters in the plasma because the iota profile is strongly deformed in the inner region and it falls below $\rlap{-} \iota = 1/3$ (Fig. 8 b). 

\begin{figure}[h]
\centering
\includegraphics[width=0.35\textwidth]{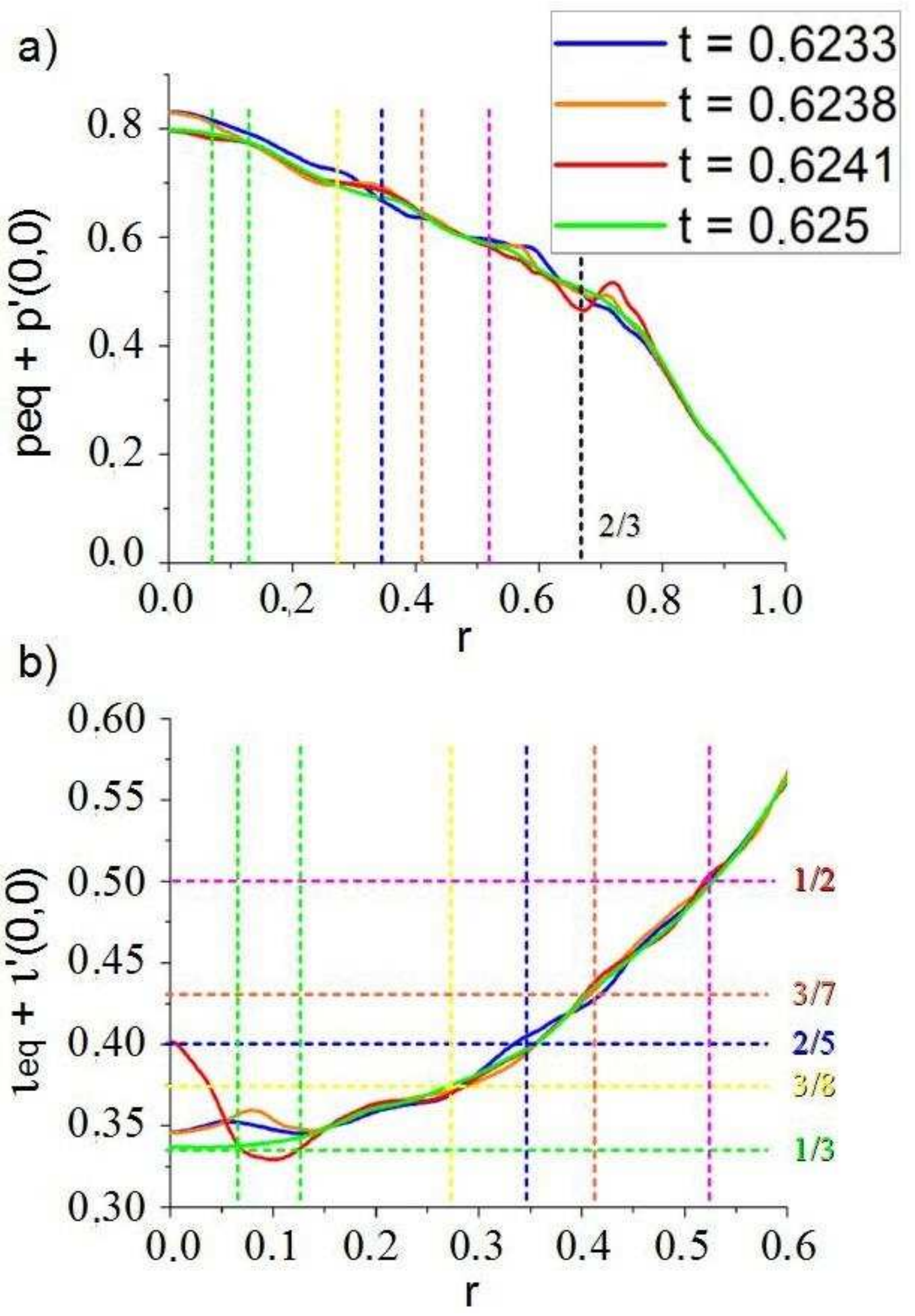}
\caption{Pressure (a) and rotational transform (b) profiles during the main event of the $S = 2.5 \times 10^5$ simulation.}
\end{figure}

The main event in the high $S$ simulation is triggered at $t = 0.6191$ s. The main deformation in the pressure profile is driven by the mode $1/2$ in the middle plasma but it is weaker than in the low $S$ simulation, Fig. 9 a. There are several isolated deformations between the middle and outer plasma driven by the modes $3/5$, $2/3$, $3/4$ and $1/1$, but not in the inner plasma. The iota profile is deformed near the magnetic axis but it doesn't drop below $\rlap{-} \iota = 1/3$, Fig. 9 b. 

\begin{figure}[h]
\centering
\includegraphics[width=0.35\textwidth]{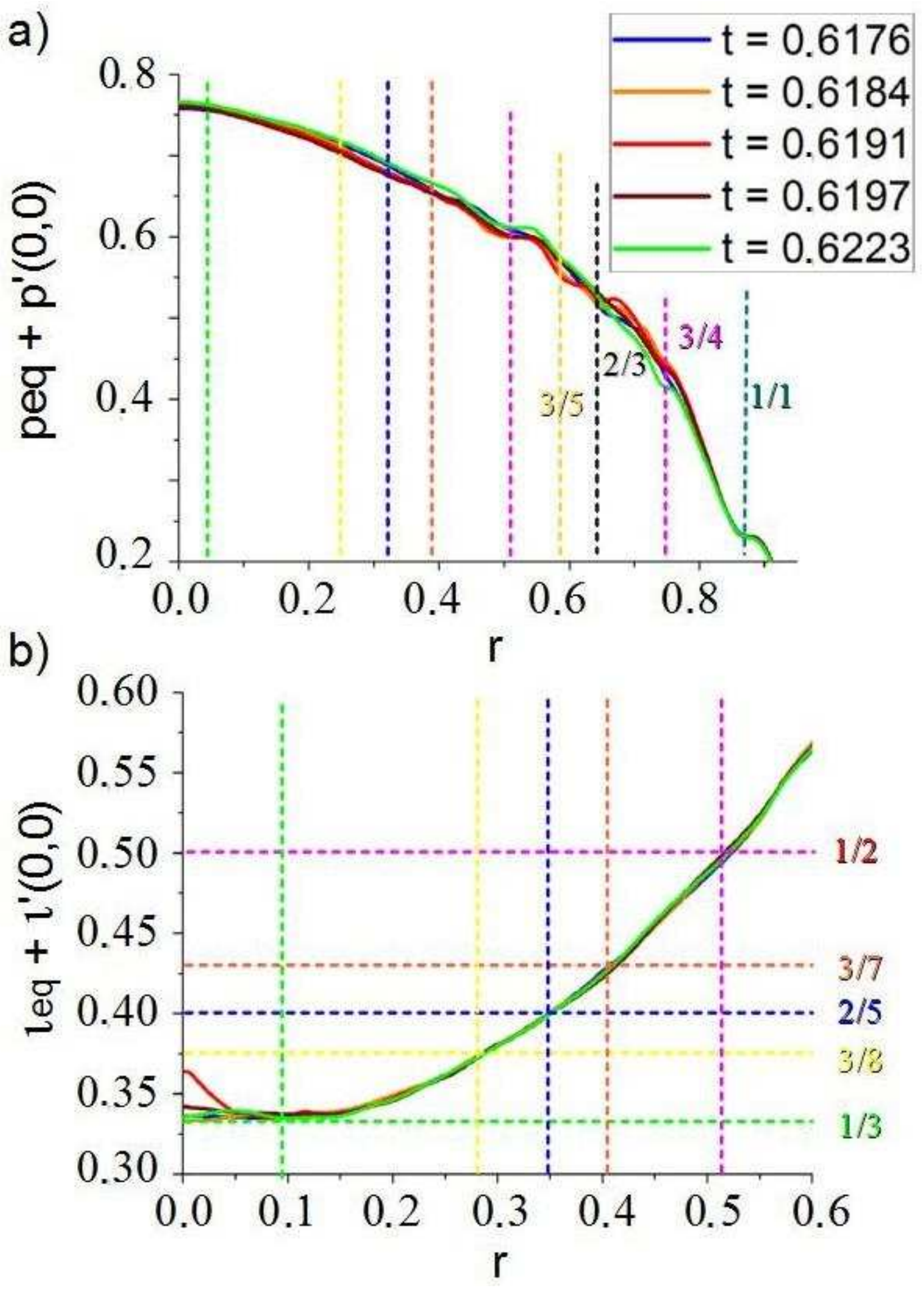}
\caption{Pressure (a) and rotational transform (b) profile during the main event of the $S = 10^6$ simulation.}
\end{figure} 

The poloidal contours of the pressure in the low $S$ case, Fig. 10, show a strong deformation of the flux surface in the middle and inner plasma. The flux surface between the middle and outer plasma are torn at $t = 0.6238$ and $0.6241$ s (yellow circle) and an amount of plasma is expelled. At $t = 0.6245$ s the flux surfaces deformation smooths in the inner and outer plasma but remains in the middle plasma.

\begin{figure}[h]
\centering
\includegraphics[width=0.3\textwidth]{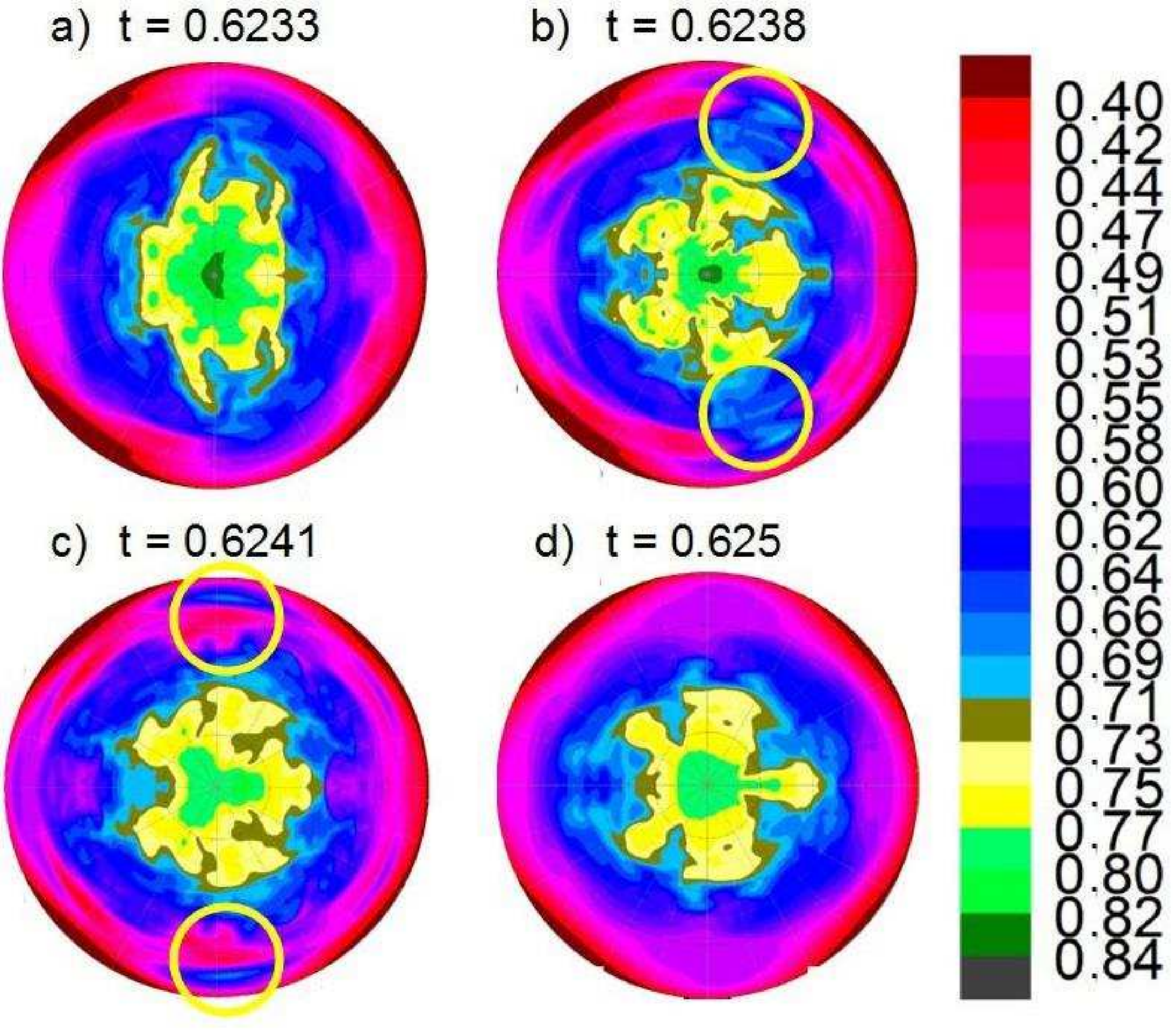}
\caption{Poloidal pressure contour during the main event of the $S = 2.5 \times 10^5$ simulation. The yellow circles show the torn flux surfaces.}
\end{figure}

The flux surfaces in the high $S$ simulation are deformed in the middle and outer plasma, Fig. 11, but the perturbation is much weaker than in the low $S$ case, specially in the inner plasma region. At $t = 0.6184$ and $0.6191$ s the flux surfaces between the middle and outer plasma are slightly torn. 

\begin{figure}[h]
\centering
\includegraphics[width=0.4\textwidth]{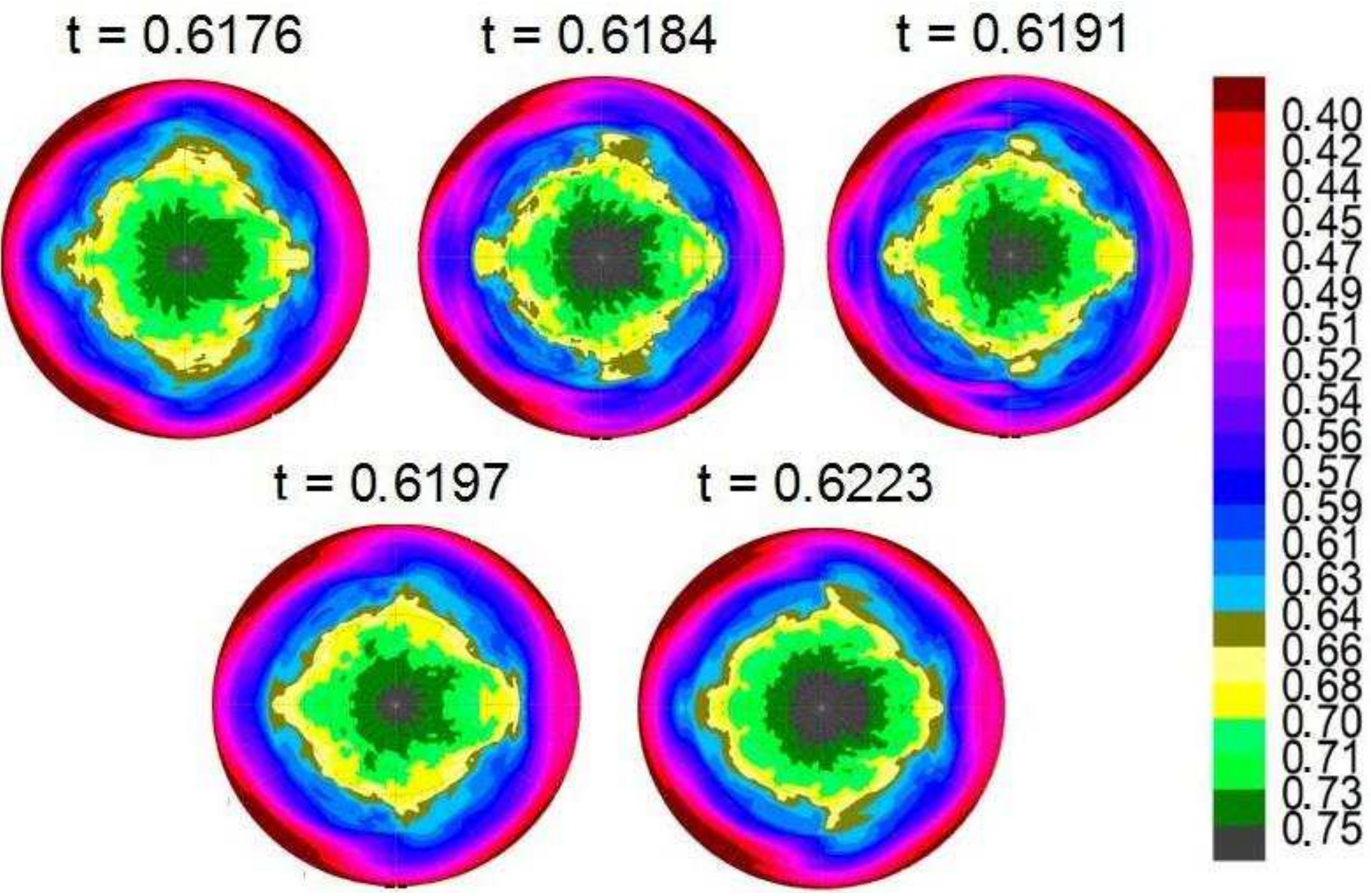}
\caption{Poloidal pressure contour during the main event of the $S = 10^6$ simulation.}
\end{figure} 

The magnetic islands overlapping in the low $S$ simulation, Fig. 12 a, is strong between the middle and the outer plasma at $t = 0.6233$ s. There are two large stochastic regions, one between the middle and the outer plasma and another between the middle and the inner plasma, Fig. 12 b. At $t = 0.6238$ s the magnetic surfaces are strongly deformed in the inner plasma and three islands appear near the magnetic axis. The outer plasma and the inner plasma are linked by a large stochastic region that almost reaches the magnetic axis. At $t = 0.6241$ s the magnetic islands overlapping decreases in the inner plasma but the $1/3$ islands still remains close to the magnetic axis. At $t = 0.6250$ s the magnetic islands overlapping keeps decreasing and the mode $1/3$ leaves the plasma. The shape of the magnetic surfaces is recovered in the inner plasma and the stochastic region between the inner and middle plasma is divided in several unlinked stochastic regions.

\begin{figure}[h]
\centering
\includegraphics[width=0.5\textwidth]{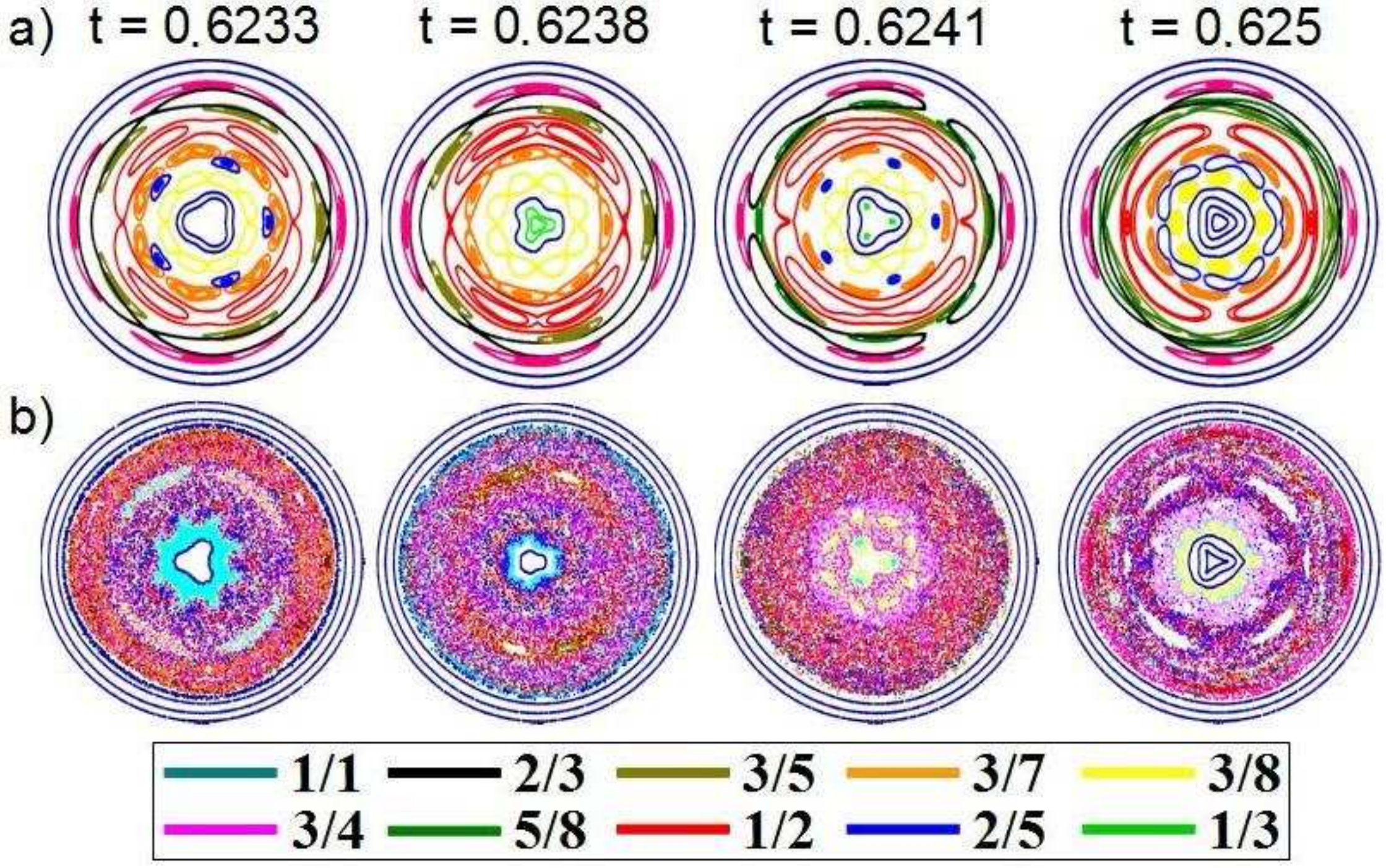}
\caption{Magnetic islands (a) and stochastic regions (b) in the main event of the $S = 2.5 \times 10^5$ simulation.}
\end{figure}

The magnetic islands overlapping in the high $S$ simulation is small at $t = 0.6176$ s, Fig. 13 a, and there are several not linked stochastic regions between the inner and the outer plasma, Fig. 13 b. The size of the magnetic islands increases between $t = 0.6184$ and $0.6197$ s and they overlap between the middle and the outer plasma but not in the inner region. The stochastic regions in the middle and outer plasma grow and they are linked at $t = 0.6191$ s, but they don't reach the inner plasma. At $t = 0.6223$ s the magnetic island width decreases and the magnetic surfaces are recovered in the middle and outer plasma. The stochastic regions are smaller and they are not linked.

\begin{figure}[h]
\centering
\includegraphics[width=0.5\textwidth]{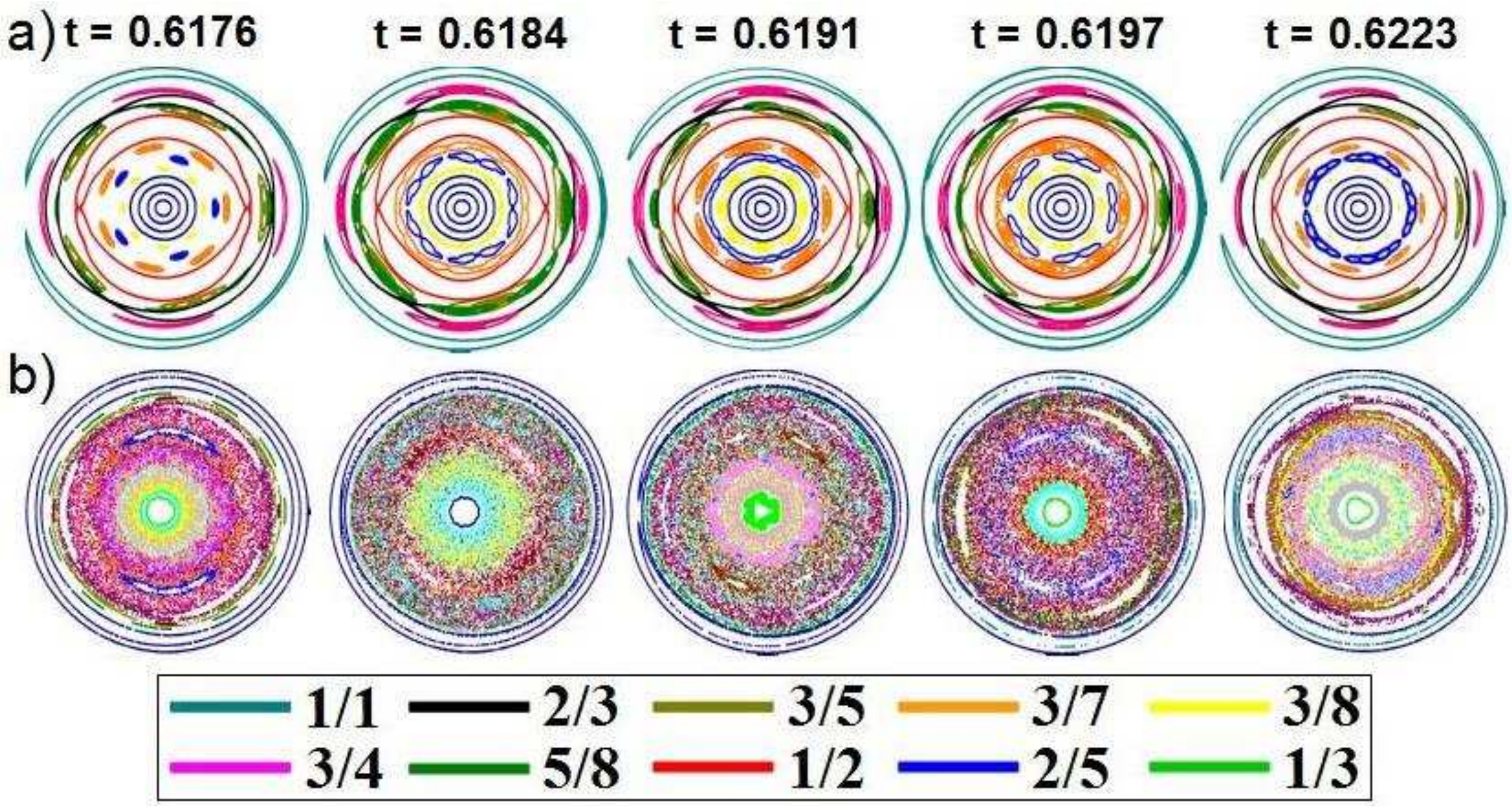}
\caption{Magnetic islands (a) and stochastic regions (b) during the main event of the $S = 10^6$ simulation.}
\end{figure} 

The averaged magnetic potential in the low $S$ simulation, Fig. 14 a, shows a magnetic well in the inner plasma and a magnetic hill between the middle and outer plasma. From $t = 0.6233$ to $0.6241$ s the magnetic well region decreases from $\rho = 0.4$ to $0.3$, therefore the plasma region unstable to interchange modes increases. The magnetic hill region is enhanced in the middle and outer plasma where the interchange modes can be destabilized easily, specially between $\rho = 0.5$ and $0.6$.  At $t = 0.625$ s the magnetic well reaches again $\rho = 0.4$ and the magnetic hill is weakened in the middle and outer plasma. The main unstable regions are located around $\rho = 0.5$ and from $\rho = 0.65$ to the plasma periphery. In the high $S$ case, Fig. 14 b, the magnetic well only changes slightly, oscillating between $\rho = 0.4$-$0.45$. The magnetic hill region is not enhanced and the local maximums are around the modes $1/2$, $3/4$ and $1/1$. These results point out that the effect of the main events in the plasma equilibria is weaker in the high $S$ simulation than in the low $S$ case, therefore the plasma stability is more robust in the high $S$ simulation.

\begin{figure}[h]
\centering
\includegraphics[width=0.4\textwidth]{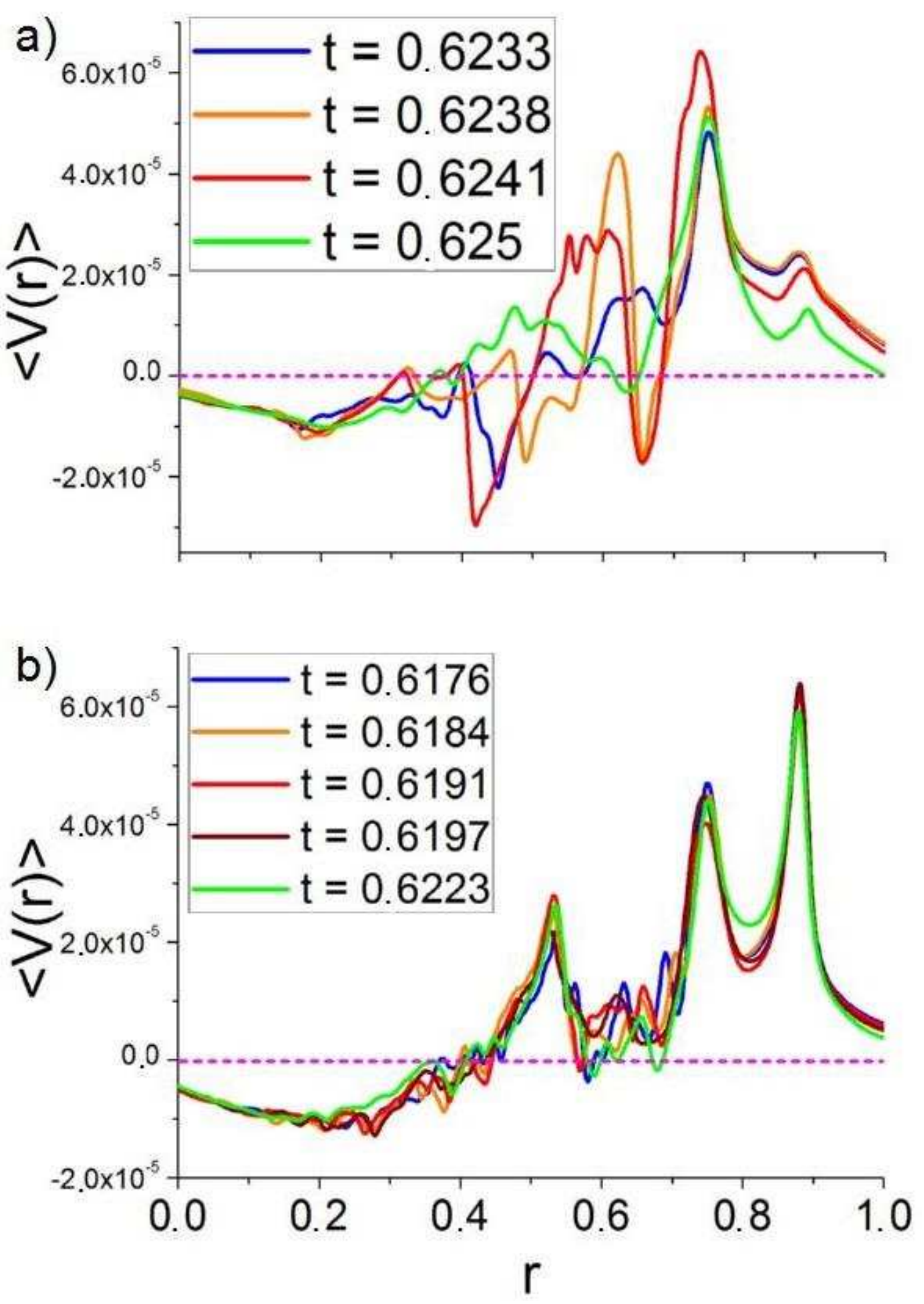}
\caption{Magnetic potential during the main events of the $S = 2.5 \times 10^5$ (a) and $S = 10^6$ (b) simulations.}
\end{figure}

The features of the main event in the low $S$ simulation shows a system relaxation in the hard MHD regime similar to a plasma collapse. It is a global relaxation where the flux and magnetic surfaces are strongly destabilized, with wide stochastic regions due to the large overlapping between magnetic islands. The main event in the high $S$ case is a relaxation in the soft MHD regime. The distortion of the flux and magnetic surfaces is local and weaker compared with the low $S$ case. The magnetic island overlaping is smaller as well as the size of the stochastic regions, similar to the $1/2$ sawtooth like events observed in LHD. 

\section{Summary and discussion \label{sec:conclusions}}

The study conclusions point out that the feedback effect between the magnetic turbulence and the pressure gradients is important to understand the MHD stability of the system and the transition between the MHD regimes. The magnetic turbulence perturbs the flux and magnetic surfaces enhancing the pressure gradients near the rational surfaces and the destabilizing effect of the unstable modes. If the destabilizing effect is large enough, broad magnetic island appear around the rational surfaces leading to a flux rearrangement that reinforces the turbulence and the pressure gradients in a feedback process. The extreme case is reached if the magnetic islands are strongly overlapped. Large regions of stochastic magnetic field will cover part of the plasma, leading to a system relaxation in the hard MHD limit.

The different Lundquist numbers of the simulations modify the characteristics of the magnetic reconnections that take place during the system evolution, leading to plasmas with different stability properties. The magnetic reconnection rate in the low $S$ simulation is larger than in the high $S$ case and it is a more efficient source of magnetic turbulence, therefore the plasma is more unstable.

The simulation with low Lundquist number describes a plasma where the magnetic turbulence and the pressure gradients increase in a feedback process until the critical value of the soft/hard transition is reached and a plasma collapse is driven. During the main event the destabilizing effect of the mode $1/2$ strongly disturbs the flux and magnetic surfaces in the inner plasma. The iota profile drops below $1/3$ and three islands appear near the magnetic axis. There is a pressure profile flattening in the inner plasma and a large profile deformation in the middle plasma driven by the mode $1/2$. The flux surfaces in the middle plasma are torn and an amount of plasma is expelled, follow by a large drop of the system energy. The magnetic well region decreases in the inner plasma and the magnetic hill is enhanced in the middle and outer plasma. There is a step drop of the pressure gradient and an abrupt increase of the magnetic turbulence.

The high $S$ simulation remains in the soft MHD regime because the magnetic turbulence decreases and the pressure gradient don't build up during the system evolution. In the main event there is a pressure profile flattening in the middle plasma driven by the mode $1/2$, but its destabilizing effect doesn't reach the inner plasma region where the profile deformation is weak. The magnetic islands width is small and the overlapping is weak between the middle and inner plasma regions. The magnetic well and hill regions keep almost unchanged. The magnetic surfaces in the inner plasma are only slightly distorted and the stochastic region in the middle plasma doesn't reach the inner plasma. The pressure gradient and the magnetic turbulence are below the critical value to drive the transition to the hard MHD limit.

The simulation with $S = 10^6$ describes in first approximation the evolution of a plasma with characteristics similar to a plasma in the middle region of a LHD for an inward-shifted configuration with a $\beta_{0} \approx 1.184$ $\%$ and line averaged electron density of $\bar{n_{e}} > 10^{20} m^{-3}$. This LHD operation with relative low $\beta$ and high density leads to a plasma with collisional features in the middle plasma region. The Lundquist number drops to a value close to $S = 10^6$ and the system remains in the slow reconnection regime, marginal unstable to the plasmoid instability. In this regime the magnetic reconnection is less efficient like a source of magnetic turbulence because the reconnection rate is slower. The destabilizing effect of the magnetic turbulence over the flux and magnetic surfaces decrease and the plasma is less unstable because the pressure gradients are weakened in a counter feedback process, leading to the reduction of the magnetic islands width and the magnetic island overlapping as well as the size of the stochastic regions. The hard MHD limit relaxations are not triggered because the magnetic turbulence and the pressure gradient don't reach the critical value to drive the soft/hard transition and the system remains in the soft MHD regime.

We propose that there is not evidence of a plasma collapse driven by the mode $1/2$ in LHD because the Lundquist number of the plasma in the middle region is large enough to keep the system in the soft MHD limit. The system remains in the slow reconnection regime and the plasmoid instability is only marginally unstable, therefore the relaxations triggered are the $1/2$ sawtooth like events. Advance operations scenarios in LHD inward-shifted configurations must remain in the soft LHD limit to avoid strong relaxation events or a plasma collapse. These operation conditions are achieved avoiding plasmas with low Lundquist numbers in the inner and middle plasma region, because the reconnection rate will be high and a very efficient source of magnetic turbulence, leading to an unstable plasma due to the feedback effect with the pressure gradients. At the same time, if the Lundquist number is high in the middle plasma region the system will evolve from the slow to the fast reconnection regime and the plasma will be unstable to the plasmoid instability. In the slow reconnection regime, the plasma stability improves if the system remains in the soft MHD limit. The pressure gradient and magnetic turbulence must be below the critical value to trigger the transition to the hard MHD limit. 

\begin{acknowledgments}
This study is supported by the SHOCK European project and the group LESIA of the Observatory Paris-Meudon. The authors are very grateful to L. Garcia for letting us use the FAR3D code and his collaboration in the developing of the present manuscript diagnostics. Thanks to F. Pantellini for discussion.
\end{acknowledgments}

\end{document}